\documentclass[prl,aps,twocolumn,10pt,superscriptaddress]{revtex4-2} 
\usepackage{amsmath,graphicx,color,gensymb,bm}
\usepackage{dcolumn}
\usepackage{amsthm}     
\usepackage{amsfonts}
\usepackage{algorithmic}
\usepackage{enumerate}
\usepackage{latexsym}
\usepackage{amssymb}
\usepackage{multirow}
\usepackage{array}
\usepackage{diagbox}
\usepackage{subfigure}
\usepackage[colorlinks=true,citecolor=blue,linkcolor=blue]{hyperref}

\begin{document}

\preprint{APS/123-QED}

\title{Stripe-Ordered Altermagnetism Emerging from Correlation-Driven Spin-Density-Wave Instability}

\author{Zenghui Fan}
\affiliation{School of Physics and Astronomy, Beijing Normal University, and Key Laboratory of Multiscale Spin Physics (Beijing Normal University), Ministry of Education, Beijing 100875, China}

\author{Jingyao Meng}
\affiliation{School of Physics and Astronomy, Beijing Normal University, and Key Laboratory of Multiscale Spin Physics (Beijing Normal University), Ministry of Education, Beijing 100875, China}

\author{Tianxing Ma}
\email{txma@bnu.edu.cn}
\affiliation{School of Physics and Astronomy, Beijing Normal University, and Key Laboratory of Multiscale Spin Physics (Beijing Normal University), Ministry of Education, Beijing 100875, China}

\begin{abstract}
Altermagnetism is conventionally identified within the paradigm of collinear antiferromagnets. Its potential realization within other spin instabilities, such as a spin-density wave (SDW), remains a fundamentally compelling open question. Here, we combine Hartree-Fock mean-field and unbiased determinant quantum Monte Carlo methods to investigate a minimal Hubbard model relevant to iron pnictides. We reveal a novel $d_{xy}$-wave stripe-ordered altermagnetic (SOAM) insulating phase driven fundamentally by the correlation-induced $(\pi,0)$ SDW instability. Within this phase, an introduced uniaxial staggered electric potential alters the underlying symmetry: it breaks the original combined time-reversal and spatial translation symmetry ($T_{d}\mathcal{T}$) and retains a combined time-reversal and mirror invariance ($M\mathcal{T}$), thereby unlocking the pronounced nonrelativistic spin splitting. Crucially, the exact finite-size scaling from our determinant quantum Monte Carlo simulations confirms that this correlation-driven SOAM phase stably survives at accessible finite temperatures. Our study pushes the frontier of altermagnetism beyond the conventional antiferromagnetic paradigm into the realm of SDW instability, advancing the fundamental understanding of altermagnetism in strongly correlated electron systems.
\end{abstract}
\maketitle

 
\noindent
\underline{\it Introduction}---
Altermagnetism (AM) has recently emerged as a fundamental magnetism beyond conventional ferromagnetism and antiferromagnetism (AFM), uniquely combining a vanishing net magnetization with macroscopic spin splitting, notably in the absence of spin-orbit coupling~\cite{JPSJ.88.123702, PhysRevX.12.031042,PhysRevX.12.040501}. Unlike traditional antiferromagnets, altermagnets break the combined parity-time symmetry while preserving specific combined symmetries of time-reversal and crystal rotation or mirror operations~\cite{gomonay2024structure,ma2021multifunctional,PhysRevLett.134.106801,PhysRevLett.132.176702,ji2025neel}. This unique symmetry breaking induces a nonrelativistic, momentum-dependent spin splitting in the electronic band structure, which in turn triggers a series of nontrivial transport, magnetoelectric, and optical properties~\cite{vsmejkal2022anomalous,PhysRevLett.133.086503,devita2026robust,sicheler2025optical}. Therefore, it sparks immense interest for both fundamental physics and spintronic applications~\cite{PhysRevX.12.040002,adfm.202409327,song2025altermagnets}. Currently, the studies for AM mainly concentrate on explicitly introducing specific crystallographic or orbital-hopping anisotropies~\cite{ji2025neel,v38b-5by1,PhysRevLett.132.236701,PhysRevB.111.L020401,PhysRevLett.132.263402,s31h-hk2v}. However, incorporating such anisotropies usually introduces frustrations and destabilizes the correlation-driven AFM, particularly in unbiased numerical calculations of Hubbard models~\cite{PhysRevB.35.3359}. Consequently, whether AM can be spontaneously driven by electron correlations remains a compelling open question~\cite{PhysRevB.75.115103}.

Conventionally, AM is generally regarded as a collinear AFM protected by specific lattice symmetries, such as combined symmetry of time-reversal and crystal rotations, which connect opposite spin sublattices. Therefore, it has been widely expected and experimentally realized in numerous established conventional antiferromagnets. For instance, $g$-wave AM band splitting has been definitively observed in $A$-type antiferromagnets CrSb~\cite{reimers2024direct,yang2025three,lu2025signature,advs.202406529,PhysRevLett.133.206401} and MnTe~\cite{krempasky2024altermagnetic,PhysRevLett.132.036702,PhysRevB.109.115102}. The $d$-wave AM has been reported in the AFM oxychalcogenide metals KV$_2$Se$_2$O~\cite{jiang2025metallic} and Rb$_{1-\delta}$V$_2$Te$_2$O~\cite{zhang2025crystal}. The quasi-2D oxyselenide La$_2$O$_3$Mn$_2$Se$_2$, previously confirmed as collinear AFM order via neutron diffraction~\cite{PhysRevB.82.214419}, has also been proposed as a $d$-wave altermagnet~\cite{PhysRevMaterials.9.024402,Garcia_Gassull_2026}, alongside a rapidly expanding roster of other AFM candidates~\cite{GUO2023100991,PhysRevMaterials.5.014409}. However, this raises a fundamental question: beyond the paradigm of conventional AFM, can AM be stably established within other spin instabilities, such as the spin-density wave (SDW)~\cite{science.aam7127,science.adh7691,Runyu2026}? A theoretical study has proposed SDW instabilities in preexisting AM context~\cite{llrq-1k9k}. Significantly, recent experiments demonstrate a magnetic-field tuning of an altermagnetic SDW order in the kagome magnet CsCr$_3$Sb$_5$~\cite{huang2026controlling}. These advances suggest an intimate connection between AM and SDW magnetic instabilities, thereby intensifying the urgency of this critical question.

Among the diverse family of strongly correlated systems, iron-based compounds exhibit tremendous potential for hosting AM~\cite{wang2026two,li2026uncovering,PhysRevLett.104.216405,7q2y-v4sz}. As a typical class of iron-based superconductors, iron pnictides are characterized by intrinsic anisotropic inter-orbital hoppings, with their parent compounds robustly stabilizing a SDW ground state~\cite{fernandes2014drives,PhysRevX.2.021009,PhysRevLett.110.107002,PhysRevLett.119.157001,shibauchi2014quantum,fernandes2022iron}. Specifically, this ground state manifests as a spin-stripe order driven by $(\pi,0)$ spin instability~\cite{yin2014spin}, which is tied to the Fermi surface topology comprising hole pockets at the $\Gamma$ point $(0,0)$ and electron pockets at the $X/Y$ points $(\pi,0)/(0,\pi)$. These physics provide an intriguing inspiration for possible AM arising from SDW instability.

\begin{figure}[t]
\centering
\includegraphics[width=0.48\textwidth]{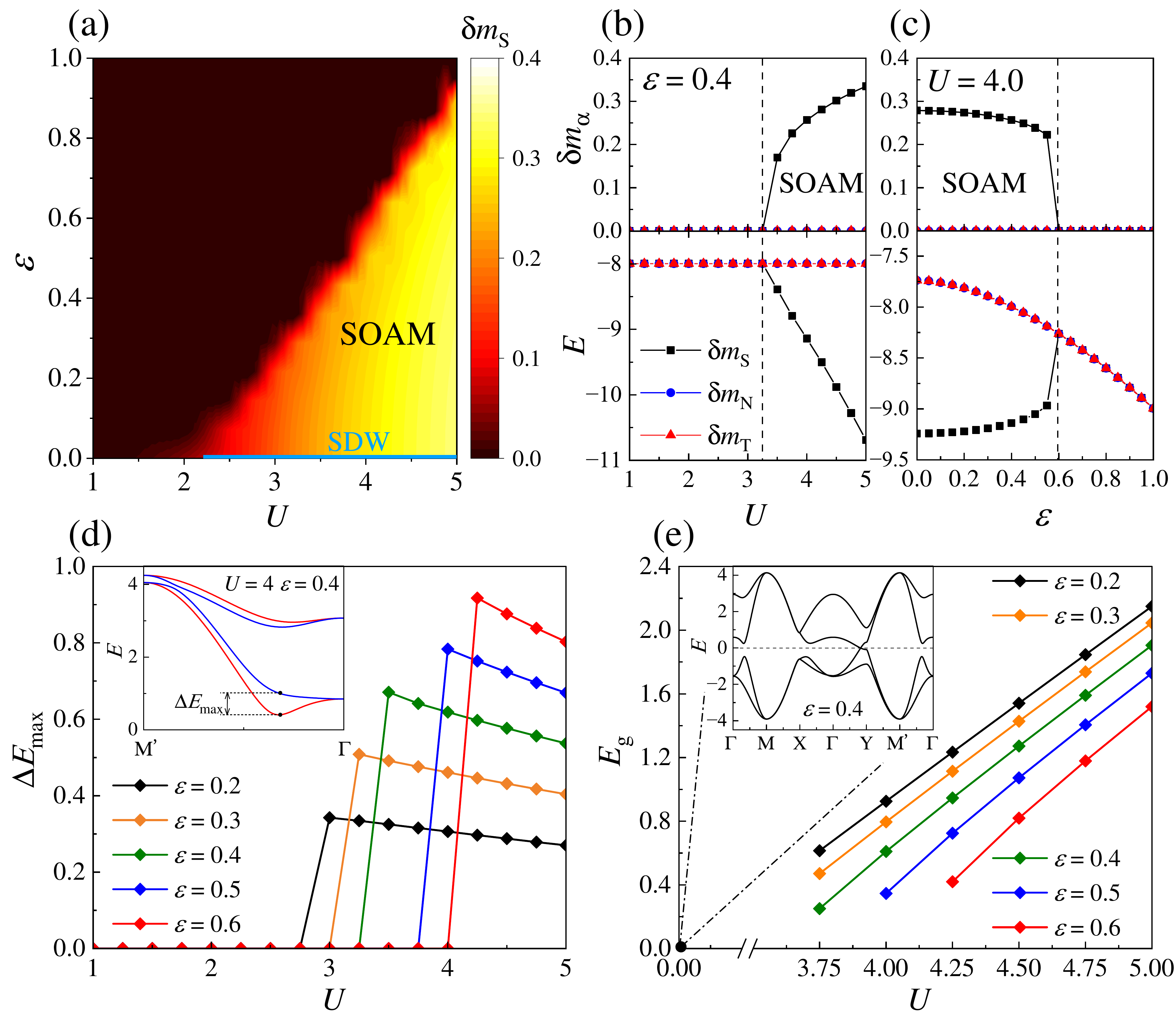}  
\caption{(a) Mean-field ground-state phase diagram at half-filling characterized by finite spin-stripe order parameter $\delta m_{\mathrm{S}}$, involving the pure SDW phase without $\varepsilon$ and SOAM phase driven by the interplay of $U$ and $\varepsilon$. Competing magnetic order parameters $\delta m_{\alpha}$ and their corresponding energy $E$ as functions of (b) $U$ and (c) $\varepsilon$ for spin stripe ($\delta m_{\mathrm{S}}$), N\'eel AFM ($\delta m_{\mathrm{N}}$), and total ferromagnetism ($\delta m_{\mathrm{T}}$). (d) Maximum spin-splitting energy difference $\Delta E_{\mathrm{max}}$ (see inset for $\Delta E_{\mathrm{max}}$) versus $U$ for various $\varepsilon$. (e) Charge gap $E_g$ within the SOAM phase as a function of $U$ for varying $\varepsilon$, with a noninteracting gapless band structure inserted for reference. }\label{fig_2}
\end{figure}

In this Letter, we combine Hartree-Fock mean-field (MF) and the unbiased determinant quantum Monte Carlo (DQMC)~\cite{PhysRevB.40.506,PhysRevLett.120.116601,Ma_2025,PhysRevB.111.045151,Meng_2025} methods to investigate a minimal Hubbard model relevant to iron pnictides. This model captures the essential Fermi surface topology and the $(\pi,0)$ SDW instability prevalent in iron pnictides. 
With a charge inhomogeneity, the correlation $U$ spontaneously drive a novel stripe-ordered AM (SOAM) emerging from the SDW instability, which integrates high anisotropy in both real-space spin texture and momentum-space spin splitting. 
The charge inhomogeneity introduced by a uniaxial staggered electric potential (USEP) $\varepsilon$ breaks the original combined time-reversal and spatial translation symmetry and preserves a combined time-reversal and mirror invariance, which unlocks a pronounced nonrelativistic $d_{xy}$-wave spin splitting. 
Despite the competing interplay between the $\varepsilon$-governed symmetry and the $U$-driven spin stripe, their remarkable coexistence establishes a broad SOAM insulating region, as shown in Fig.~\ref{fig_2}(a).
Intriguingly, our unbiased DQMC simulations confirm that the SOAM phase is established even at accessible finite temperatures. By pushing the frontier of AM beyond the conventional AFM paradigm into the realm of SDW instability, our study advances the fundamental understanding of AM in strongly correlated electron systems.

\begin{figure*}[t]
\centering
\includegraphics[width=0.8\textwidth]{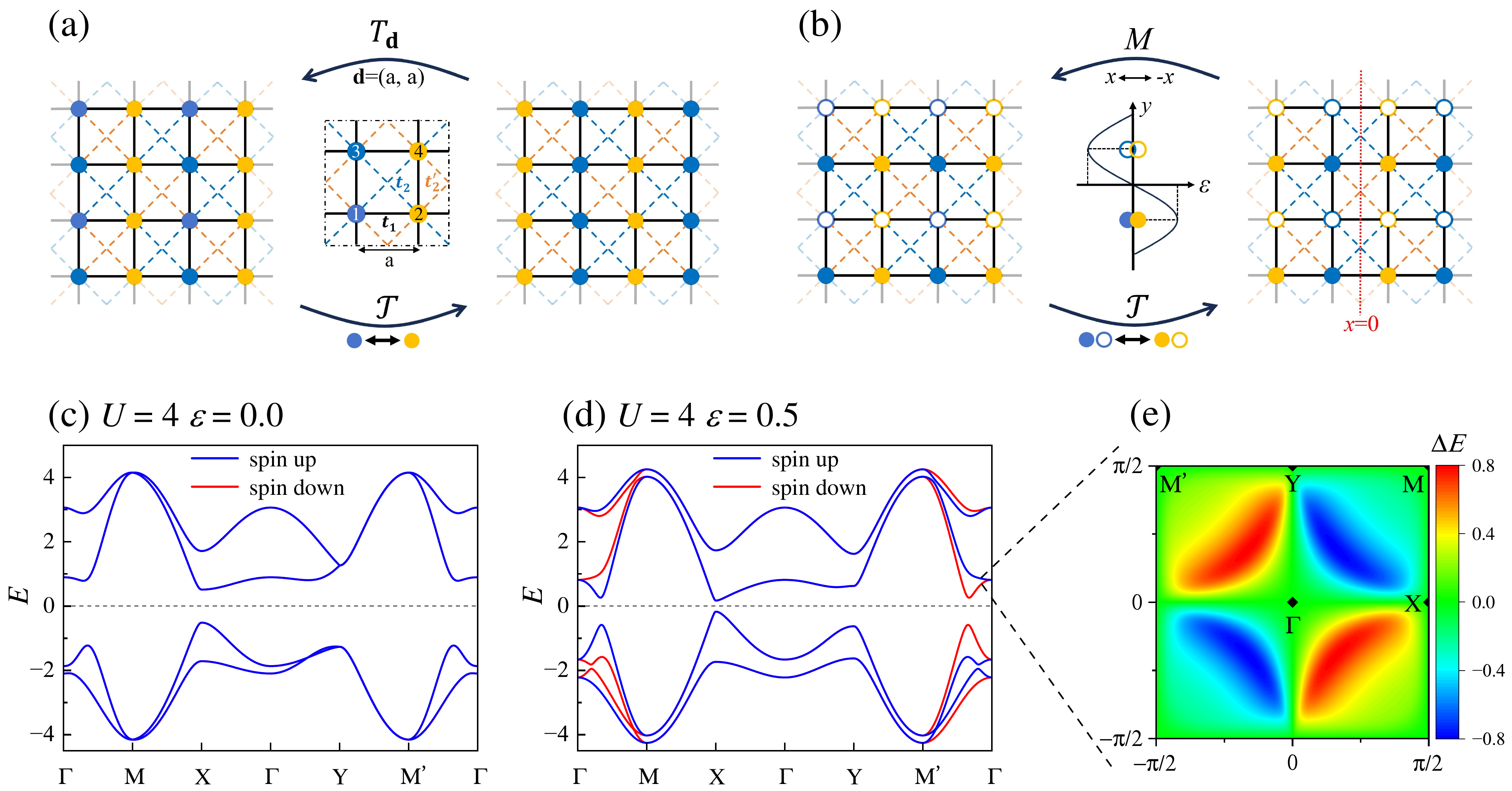}  
\caption{Real-space spin-stripe configurations (a) without and (b) with the USEP. Yellow and blue spheres represent spin-up and spin-down electrons, respectively. There is four-site magnetic unit cell (sublattice $l=1,2,3,4$ and lattice constant $a=1$) with staggered NNN hopping inserted in (a). In (b), filled and open circles denote the charge-rich and hole-rich sites induced by the USEP sketched in the inset. The pure spin stripe in (a) preserves the combined time-reversal ($\mathcal{T}$) and diagonal-translation ($T_{d}$) symmetry, while introducing the charge inhomogeneity via USEP in (b) breaks $T_{d}\mathcal{T}$ invariance, leaving the system protected by a combined $\mathcal{T}$ and mirror ($M$) symmetry. (c), (d) Corresponding mean-field band structures along high-symmetry momentum paths. The $M\mathcal{T}$ symmetry in (d) unlocks a pronounced spin splitting. (e) Momentum-space distribution of the spin-splitting energy difference $\Delta E=E_{\mathrm{up}}-E_{\mathrm{down}}$ for the lowest conduction band in (d), indicating an unconventional $d_{xy}$-wave AM. }\label{fig_1}
\end{figure*}

\noindent
\underline{\it Model and Method}---
We consider the Hubbard model defined on a square lattice with a four-sublattice basis under a USEP. The Hamiltonian is given by 
\begin{eqnarray}\label{H1}
\hat{H} &=&\sum_{{\bf i}l,{\bf j}l'\sigma} t_{{\bf i}l,{\bf j}l'} (c_{{\bf i}l\sigma}^\dagger c_{{\bf j}l'\sigma} + \text{H.c}) + \mu\sum_{{\bf i}l}n_{{\bf i}l}\nonumber\\
&&- \varepsilon\sum_{{\bf i}}(n_{{\bf i}1}+n_{{\bf i}2}-n_{{\bf i}3}-n_{{\bf i}4})\nonumber\\
&&+ U \sum_{{\bf i}l} n_{{\bf i}l\uparrow} n_{{\bf i}l\downarrow},
\end{eqnarray}
where $c_{{\bf i}l\sigma}^\dagger$ ($c_{{\bf i}l\sigma}$) creates (annihilates) an electron with spin $\sigma \in \{\uparrow,\downarrow\}$ on sublattice $l \in \{1, 2, 3, 4\}$ in unit cell ${\bf i}$, and $n_{{\bf i}l\sigma}=c_{{\bf i}l\sigma}^\dagger c_{{\bf i}l\sigma}$ is the occupation number operator with $n_{{\bf i}l} = n_{{\bf i}l\uparrow} + n_{{\bf i}l\downarrow}$. The hopping matrix elements $t_{{\bf i}l,{\bf j}l'}$ involves a uniform nearest-neighbor (NN) hopping $t_1$ and two distinct next-nearest-neighbor (NNN) hoppings, $t_2$ and $t_2'$. As sketched in Fig.~\ref{fig_1}(a), these NNN hoppings alternate with $t_2$ and $t_2'$ as follows: along the $(1,1)$/$(1,-1)$ directions, the hopping element is $t_2$/$t_2'$ for sublattices $l=1$ and $4$, and $t_2'$/$t_2$ for sublattices $l=2$ and $3$. The parameter $\varepsilon$ quantifies the USEP, which introduces charge inhomogeneity on sublattices, while $\mu$ is the chemical potential and $U$ denotes the on-site Hubbard repulsion.

Our system originates from a classical iron-based model with $S_4$ symmetry~\cite{PhysRevX.2.021009}, which effectively abstracts the underlying electronic and magnetic structure. 
By setting $t_1 = 0.3$, $t_2 = 1.4$, and $t_2' = -0.6$, we capture the essential Fermi surface structure and magnetic fluctuations prevalent in typical iron pnictides~\cite{fernandes2014drives,PhysRevX.2.021009,PhysRevLett.110.107002,PhysRevLett.119.157001,shibauchi2014quantum,fernandes2022iron} (see {\color{blue}Supplementary Materials (SM), Sec. S1} for details). 
Focusing on the half-filled case, we successively employ the Hartree-Fock MF method and unbiased DQMC simulations to comprehensively characterize a SOAM phase emerging from the SDW instability.

In order to study the magnetic instablities of this system, we define a set of collinear MF order parameters within the $2\times2$ unit cell. In terms of the local spin operator $S^{z}_{{\bf i}l}=n_{{\bf i}l\uparrow}-n_{{\bf i}l\downarrow}$ and the total number of unit cells $N_c$, the conventional N\'eel antiferromagnetic and total ferromagnetic parameters are given by $\delta m_{N/T}=(1/8N_c)\sum_{\bf i}\langle S^{z}_{{\bf i}1}\mp S^{z}_{{\bf i}2}\mp S^{z}_{{\bf i}3}+ S^{z}_{{\bf i}4}\rangle$. Additionally, we introduce the spin-stripe order parameter $\delta m_{S}=(1/8N_c)\sum_{\bf i}\langle S^{z}_{{\bf i}1}- S^{z}_{{\bf i}2}+ S^{z}_{{\bf i}3}- S^{z}_{{\bf i}4}\rangle$, which properly captures the $(\pi,0)$ magnetic instability. We utilize $\delta m_{S}$ to illustrate the subsequent Hartree-Fock analysis method.

Considering the half-filling condition, $n = \frac{1}{N_c}\sum_{{\bf i}l}\langle n_{{\bf i}l}\rangle = 4$, the local electron occupation for the spin-stripe configuration is parameterized as
\begin{equation}\label{dm}
\langle n_{{\bf i}l\sigma} \rangle = \frac{1}{2} + (-1)^{l+\sigma}\delta m_{S},
\end{equation}
where $\sigma$ is associated with values $1, 2$ for $\uparrow, \downarrow$, respectively. Applying a standard Hartree-Fock decoupling to the interaction term, an effective interaction $H_{U}=-U\sum_{l\sigma}(-1)^{l+\sigma}\delta m_{S}n_{{\bf k}l\sigma}$ is obtained after a Fourier transformation. Hence, the effective Hamiltonian in momentum space can be expressed as,
\begin{equation}\label{HF}
\hat{H}^{\text{HF}} = \sum_{\bf k} \Psi_{\bf k}^{\dagger} 
\begin{bmatrix} 
H_{{\bf k}\uparrow} & 0 \\ 
0 & H_{{\bf k}\downarrow} 
\end{bmatrix} 
\Psi_{\bf k},
\end{equation}
defined in the basis $\Psi_{\bf k}^\dagger = (\Psi_{{\bf k}\uparrow}^\dagger, \Psi_{{\bf k}\downarrow}^\dagger)$ with $\Psi_{{\bf k}\sigma}^\dagger = (c_{{\bf k}1\sigma}^\dagger, c_{{\bf k}2\sigma}^\dagger, c_{{\bf k}3\sigma}^\dagger, c_{{\bf k}4\sigma}^\dagger)$. 
The off-diagonal matrix elements of the $4\times 4$ blocks $H_{{\bf k}\sigma}$ are determined by the NN and NNN hopping elements between sublattices, whereas the effects of $U$ and $\varepsilon$ are confined to the diagonal elements, yielding
\begin{equation}\label{H_diag}
H_{{\bf k}\sigma}(l,l) = \mu - (-1)^{l+\sigma}\delta m_{S}U \mp \varepsilon,
\end{equation}
with $-\varepsilon$ $(+\varepsilon)$ for $l=1, 2$ $(3, 4)$. Analogous decoupling analysis are applied for order parameters $\delta m_{N}$ and $\delta m_{T}$. We solve their MF equations by self-consistently determining these order parameters with corresponding $\mu$ in the zero-temperature ground state (see {\color{blue}SM, Sec. S2}).

To characterize accurately the SOAM at finite temperatures, we employ the unbiased DQMC method beyond the MF approximation. However, the inclusion of NNN hoppings breaks the particle-hole symmetry, thus driving the system away from half-filling and introducing the sign problem (see {\color{blue}SM, Sec. S3}). Within tolerable sign-problem regime, we tune $\mu$ to achieve the half-filled electron occupation. 
To capture the magnetic correlations under the charge inhomogeneity ($\varepsilon\neq0$), we compute the momentum-resolved spin structure factor,
\begin{equation}\label{Ss}
S({\bf k}) = \frac{1}{N_c} \sum_{{\bf i}l,{\bf j}l'} \mathrm{e}^{-\mathrm{i}{\bf k}\cdot({\bf R}_{{\bf i}l}-{\bf R}_{{\bf j}l'})} \langle S_{{\bf i}l}^z S_{{\bf j}l'}^z \rangle.
\end{equation}
The spin-stripe instability of MF parameter $\delta m_S$ manifests as the magnitude of $S(\pi,0)$, which is defined as spin-stripe structure factor $O_s \equiv S(\pi,0)$.

\noindent
\underline{\it Results and discussion}---
Our system follows an iron-based model~\cite{PhysRevX.2.021009,PhysRevLett.110.107002}, which captures the essential Fermi surface topology and the characteristic $(\pi,0)$ SDW instability. As illustrated in Fig.~\ref{fig_1}(a), the basic spin-stripe magnetic unit is defined on a $2\times2$ plaquette with distinct NNN hoppings, alternating between $t_{2}$ and $t_{2}'$. While this collinear spin stripe configuration ensures a vanishing net magnetization, it remains invariant under the combined operation of time-reversal ($\mathcal{T}$) and a diagonal translation ($T_{d}$). This $T_{d}\mathcal{T}$ symmetry strictly prohibits any AM spin splitting. Consequently, although a substantial electron correlation $U$ opens a SDW gap, the spin bands remain completely degenerate in Fig.~\ref{fig_1}(c). 

A fundamental question thus arises: can nonrelativistic spin splitting emerge within such a spin-stripe background, beyond the conventional AFM paradigm? To explore this, we introduce a sublattice-resolved charge inhomogeneity via the USEP $\varepsilon$ to merely alter the underlying symmetry, which generates a charge density modulation that differentiates the sublattices into charge-rich and hole-rich sites, as sketched in Fig.~\ref{fig_1}(b). As a result, the system breaks the original $T_{d}\mathcal{T}$ symmetry and remains invariant under a combined $M\mathcal{T}$ operation, where $M$ denotes a mirror reflection. This new symmetry connects opposite spin sublattices which unlocks a nonrelativistic spin splitting in specific momentum regions, as clearly resolved in Fig.~\ref{fig_1}(d). 

Coupled with the zero net magnetization, this definitely signals the establishment of a novel SOAM arising from the SDW instability. To visualize this, we map the spin-splitting energy difference, $\Delta E = E_{\mathrm{up}}-E_{\mathrm{down}}$, of the lowest conduction band across the first Brillouin zone in Fig.~\ref{fig_1}(e). The spin-band splitting is concentrated along the Brillouin zone diagonals, with symmetry-protected nodes along the paths from $(0,0)$ to $(0,\pm\pi/2)$ and $(\pm\pi/2,0)$. This profile firmly establishes an unconventional $d_{xy}$-wave SOAM, which integrates high anisotropy in both real-space spin texture and momentum-space spin splitting. It demonstrates the significant connection between AM and correlated SDW instability in strongly correlated electron systems.

To explore the magnetic instability, we perform Hartree-Fock MF calculations allowing for all possible collinear spin configurations within the $2\times2$ unit cell, including the spin-stripe ($\delta m_S$), N\'eel antiferromagnetic ($\delta m_N$), and total ferromagnetic ($\delta m_T$) order parameters. Self-consistent solutions reveal that the ground state is robustly dominated by the correlated SDW instability, characterized by a finite $\delta m_S$. The resulting half-filled ground-state phase diagram under the interplay between $U$ and $\varepsilon$ is summarized in Fig.~\ref{fig_2}(a).
At $\varepsilon=0$, the preserved $T_{d}\mathcal{T}$ symmetry protects a pure correlated SDW state with completely degenerate spin bands. Under the charge inhomogeneity ($\varepsilon>0$), $U$ drives the system into the SOAM phase with $\varepsilon$-governed $M\mathcal{T}$ symmetry unlocking momentum-space spin splitting. However, increasing $\varepsilon$ suppresses the critical $U_c$ and tends to destroy this phase. Strikingly, despite the competition between the $\varepsilon$-governed symmetry and the $U$-driven spin stripe, their interplay stabilizes a broad SOAM regime where spin splitting and stripe order robustly coexist.

The energetic competition between these distinct magnetic orders is detailed in Figs.~\ref{fig_2}(b) and (c). For a fixed $\varepsilon=0.4$, a continuous transition into the SOAM phase occurs at $U_c \approx 3.25$, characterized by a sharp onset of $\delta m_S$ and a corresponding substantial reduction in the ground-state energy. By contrast, both $\delta m_N$ and $\delta m_T$ remain strictly zero across the entire parameter space, confirming the energetic preference for the spin-stripe instability. Furthermore, at a fixed $U=4.0$, a sufficient $\varepsilon_c \approx 0.6$ fully suppresses the spin-stripe configuration, thereby driving a quantum phase transition out of the SOAM phase into a nonmagnetic state.

To quantitatively capture the AM response, we track the maximum spin-splitting energy difference $\Delta E_{\mathrm{max}}$ of the lowest conduction band, as highlighted in the inset of Fig.~\ref{fig_2}(d). At $U < U_c$, the spin bands remain strictly degenerate regardless of $\varepsilon$. Interestingly, the $U_c$ abruptly drives the spin splitting due to spontaneous establishment of the spin-stripe order. The $\Delta E_{\mathrm{max}}$ is governed by the delicate interplay between $\varepsilon$ and $U$. Because the spin splitting relies on the $M\mathcal{T}$ symmetry contributed by USEP, a larger $\varepsilon$ promotes the charge inhomogeneity and monotonically enhances $\Delta E_{\mathrm{max}}$. Conversely, increasing $U$ slightly suppresses the spin splitting. This behavior is due to the fact that $U$ favors single-electron occupation and thus resists the $\varepsilon$-introduced charge inhomogeneity, driving the system back toward the $T_{d}\mathcal{T}$-symmetric limit. This contrasting response underscores the subtle physical balance between $U$-driven spin stripe and $\varepsilon$-governed symmetry is required to host this novel SOAM phase.
Moreover, the energy reduction caused by the establishment of the SOAM phase is also reflected in the charge transport properties. While the noninteracting system is gapless in inset of Fig.~\ref{fig_2}(e), the establishment of spin stripe order opens a charge gap, rendering the SOAM phase inherently insulating. As shown in Fig.~\ref{fig_2}(e), within the SOAM regime, the gap scales linearly with $U$ but is suppressed by $\varepsilon$, reflecting a consistent competition.

\begin{figure}[t]
\centering
\includegraphics[width=0.35\textwidth]{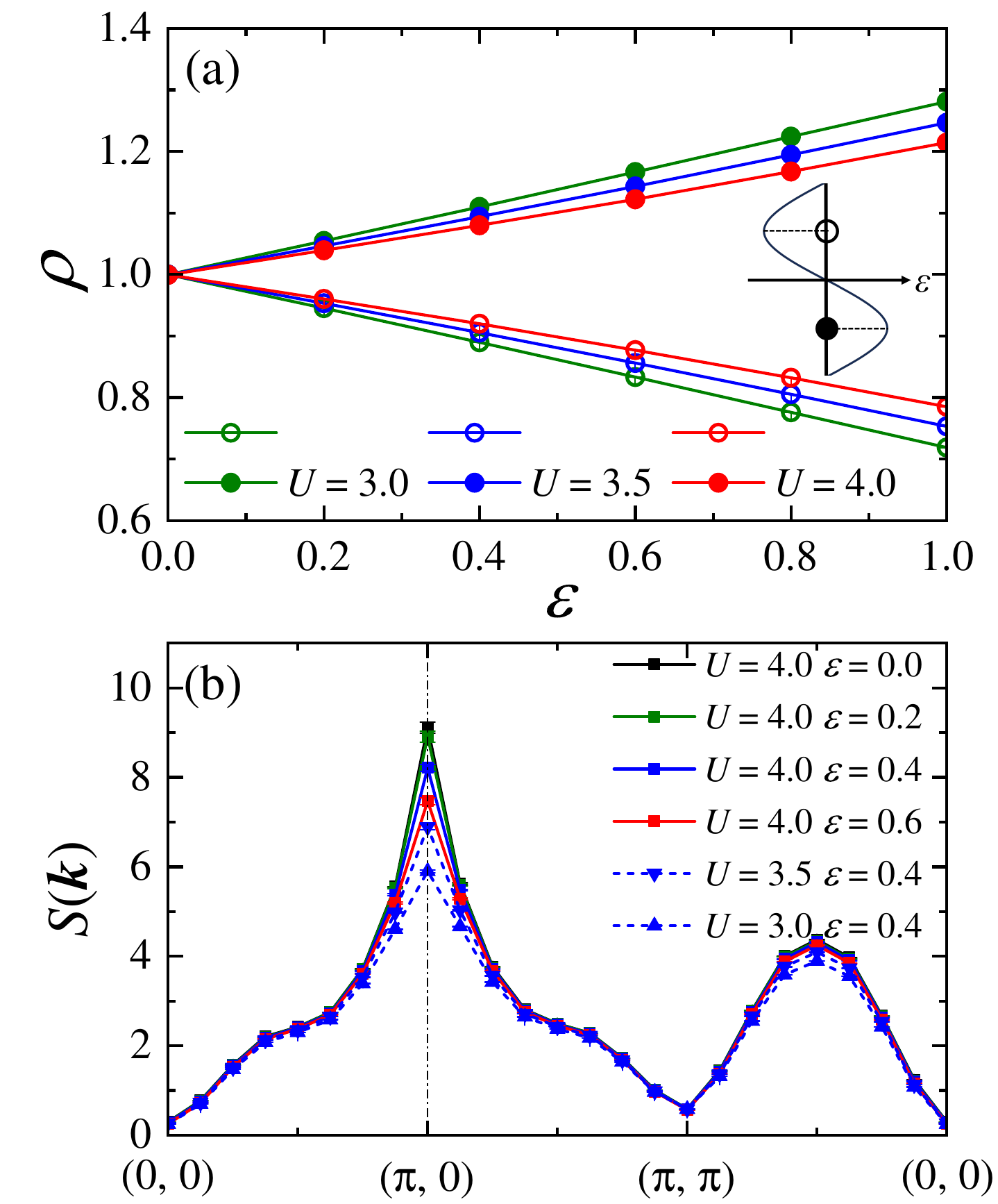}  
\caption{(a) Sublattice electron density $\rho$ on charge-rich (filled circles) and hole-rich (open circles) sites as a function of $\varepsilon$ for various $U$. (b) Response of momentum-space spin structure factor $S(\mathbf{k})$ to different $\varepsilon$ and $U$. The robust peak at $(\pi,0)$ suggests a spin-stripe order. All data are extracted from unbiased DQMC simulations at an inverse temperature $\beta=1/T=6$ with system size $L=16$.}\label{fig_3}
\end{figure}

While the MF analysis provides a clear conceptual picture, the spin splitting inherently depends on the survival of the spin-stripe order against thermal and quantum fluctuations. To establish the SOAM phase beyond the MF approximation, we perform unbiased DQMC simulations at finite temperatures. At half-filling, the USEP breaks the $T_{d}\mathcal{T}$ symmetry by differentiating sublattices into charge-rich and hole-rich sites.
To quantify this symmetry breaking, we first examine the evolution of the sublattice electron density $\rho$ as shown in Fig.~\ref{fig_3}(a). The $\varepsilon$ drives a linear deviation from half-filling ($\rho=1$) on charge-rich and hole-rich sublattices, laying the essential $M\mathcal{T}$-symmetry foundation required for AM splitting. However, $U$ favors single-electron occupation, thereby suppressing this charge inhomogeneity. 

To definitively confirm the underlying magnetic instability, we calculate the spin structure factor $S(\mathbf{k})$ in the unfolded momentum space, as shown in Fig.~\ref{fig_3}(b). The emergence of a sharp peak at $\mathbf{k} = (\pi, 0)$ provides compelling evidence for the dominant spin-stripe configuration. There reflects a competition that $\varepsilon$ dilutes the local magnetic moments and suppresses $S(\pi,0)$, whereas $U$ dramatically facilitates it and drives the spin-stripe order. These unbiased DQMC results are in excellent agreement with our MF analysis.
 
\begin{figure}[t]
\centering
\includegraphics[width=0.48\textwidth]{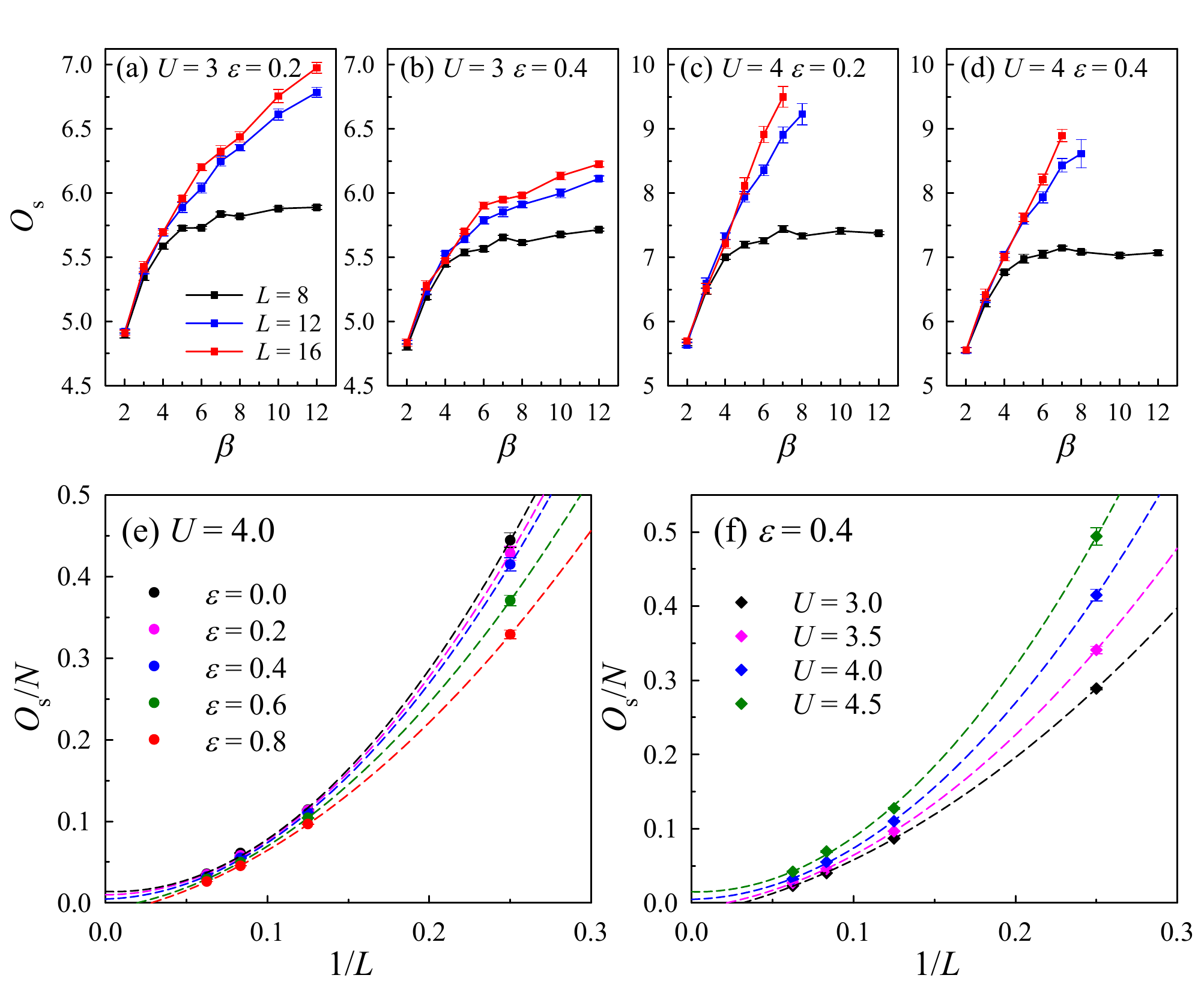}  
\caption{Finite-size scaling analysis of the spin-stripe structure factor $O_{\mathrm{s}}$ from unbiased DQMC simulations. (a)-(d) Evolution of $O_{\mathrm{s}}$ versus inverse temperature $\beta=1/T$ for various system sizes $L$ at $U=3.0$, $4.0$ and $\varepsilon=0.2$, $0.4$. The enhancement of $O_{\mathrm{s}}$ with increasing $L$ at low temperatures signifies a tendency toward spin-stripe ordering. (e), (f) Normalized structure factor $O_{\mathrm{s}}/N$ versus inverse size $1/L$ at $\beta=6$. Data are extrapolated to $1/L \to 0$ limit for (e) different $\varepsilon$ with fixed $U=4.0$ and (f) different $U$ with fixed $\varepsilon=0.4$. A finite intercept indicates the establishment of long-range order.}\label{fig_4}
\end{figure}

The most compelling evidence for the stabilization of the SOAM phase is the existence of long-range SDW order with the charge inhomogeneity. To demonstrate this, we perform a finite-size scaling analysis of the spin-stripe structure factor $O_s$ with $\mathbf{k} = (\pi,0)$ at finite temperatures, as presented in Fig.~\ref{fig_4}. We note that the inclusion of NNN hoppings breaks the particle-hole symmetry at half-filling and introduces the sign problem that becomes severe at lower temperatures $T$ and larger system sizes $L$. 
Despite this numerical constraint, the $L$-dependence of $O_s$ across various $T$ extracts crucial signatures of magnetic ordering in Figs.~\ref{fig_4}(a)-(d). In the high-$T$ regime (small $\beta$), $O_s$ is dominated by thermal fluctuations and exhibits negligible $L$-dependence. As $T$ is lowered, a pronounced finite-size effect emerges: $O_s$ increases monotonically with $L$, signaling a strong tendency toward long-range spin-stripe ordering.
This finite-size effect remains relatively weak for $U=3$ in Figs.~\ref{fig_4}(a) and (b), while it reveal a more pronounced growth in both $O_s$ and its $L$-dependence with decreasing $T$ for $U=4$ in Figs.~\ref{fig_4}(c) and (d), even though the sign problem restricts the lower-$T$ regime. The stabilization of the spin-stripe order with decreasing $T$ is also reflected by the spin susceptibility in {\color{blue} SM, Sec. S4}.
Furthermore, consistent with our MF analysis, increasing $\varepsilon$ suppresses both $O_s$ and its associated finite-size scaling.

To determine the existence of long-range SDW order, we perform a finite-size scaling analysis of $O_s$ at an accessible inverse temperature $\beta=6$ in Figs.~\ref{fig_4}(e) and (f). Employing the standard scaling hypothesis $O_s/L^2 = a + b/L + c/L^2$~\cite{PhysRevX.3.031010}, we extract $O_s/N$ in the thermodynamic limit ($1/L \to 0$). For a fixed $U=4.0$ in Fig.~\ref{fig_4}(e), the scaling curves at small $\varepsilon$ explicitly extrapolate to finite intercepts, demonstrating the presence of long-range spin-stripe order. When $\varepsilon$ is increased to $0.6$, this intercept vanishes, denoting the breaking of long-range order by charge inhomogeneity. At a fixed $\varepsilon=0.4$ in Fig.~\ref{fig_4}(f), $U=3.0$ and $3.5$ are insufficient to host the stripe phase, while a clear finite intercept emerges at $U=4.0$ and is further enhanced by $U$. 
Therefore, our results provides unbiased numerical evidence for the establishment of the SOAM phase emerging from the $U$-driven SDW instability.

\noindent
\underline{\it Conclusions}---
To explore AM arising from spin instabilities beyond the conventional AFM paradigm, we investigate a minimal Hubbard model relevant to iron pnictides using Hartree-Fock MF and DQMC methods. 
This model captures the essential Fermi surface topology and the $(\pi,0)$ SDW instability. 
By evaluating diverse MF order parameters and spin-splitting energies, we reveal that with a charge inhomogeneity introduced by $\varepsilon$, $U$ spontaneously drive a novel $d_{xy}$-wave SOAM phase emerging from the SDW instability, which integrates high anisotropy in both real-space spin texture and momentum-space spin splitting. The charge inhomogeneity alters the underlying symmetry: the original combined time-reversal and translational symmetry is broken and replaced by a combined time-reversal and mirror invariance, which unlocks a pronounced nonrelativistic spin splitting.
Although there is a competition between the $\varepsilon$-governed AM symmetry and the $U$-driven spin stripe, their remarkable coexistence stabilizes a broad SOAM insulating phase. Significantly, our exact finite-size scaling from unbiased DQMC simulations demonstrate that the SOAM phase can be stably established even at accessible finite temperatures. This provides compelling evidence for the physical realization of SOAM phase arising from SDW instability.

Our study extends AM from the conventional AFM paradigm into the realm of SDW instability, advancing the fundamental understanding of AM in strongly correlated electron systems. Admittedly, 
while our model does not capture the full microscopic complexity of real iron pnictides and focuses instead on the relevant Fermi surface topology and magnetic phenomenology, 
it nevertheless provides conceptual insights into AM in iron-based systems. 
Furthermore, given that the USEP can be readily engineered via laser interference, ultracold atomic systems in optical lattices~\cite{PhysRevLett.132.263402,RevModPhys.82.1225} emerge as a highly promising and tunable experimental platform for the realization of this novel SOAM phase.

\noindent
\underline{\it Acknowledgements:}~~
This work is supported by NSFC (12474218) and Beijing Natural Science Foundation (No. 1242022 and 1252022). 
The numerical simulations in this work were performed at the HSCC of Beijing Normal University.

\bibliography{SOAM}

\newpage
\noindent
\clearpage
\setcounter{figure}{0}\setcounter{table}{0}\setcounter{equation}{0}
\renewcommand{\thefigure}{S\arabic{figure}}
\renewcommand{\thetable}{S\arabic{table}}
\renewcommand{\theequation}{S\arabic{equation}}
\onecolumngrid

\begin{center}{\large\bfseries Supplementary Materials for ``Stripe-Ordered Altermagnetism Emerging from Correlation-Driven Spin-Density-Wave Instability''}\end{center}



\noindent
\section{S1. Electronic band structure and Fermi surface topology }
In the context of typical iron pnictides, early studies have established a parent spin-density wave state accompanied by its underlying Fermi surface topology. Specifically, the characteristic Fermi surface consists of hole pocket centered at the $(0,0)$ point and electron pocket at the $(\pi,0)$ point. Within the weak-coupling framework, the inter-pocket interactions are strongly enhanced by the spin fluctuations at the stripe wavevector $\mathbf{k}=(\pi,0)$ connecting the hole and electron pockets, thereby promoting unconventional inter-pocket superconducting pairing~\cite{Hirschfeld_2011,chubukov2012pairing,PhysRevLett.107.117001}. Our model originates from a classical $S_4$-symmetric model of iron-based superconductors~\cite{PhysRevX.2.021009,PhysRevLett.110.107002}, which crucially captures these essential physics.

\begin{figure}[h]
\centering
\includegraphics[width=0.45\textwidth]{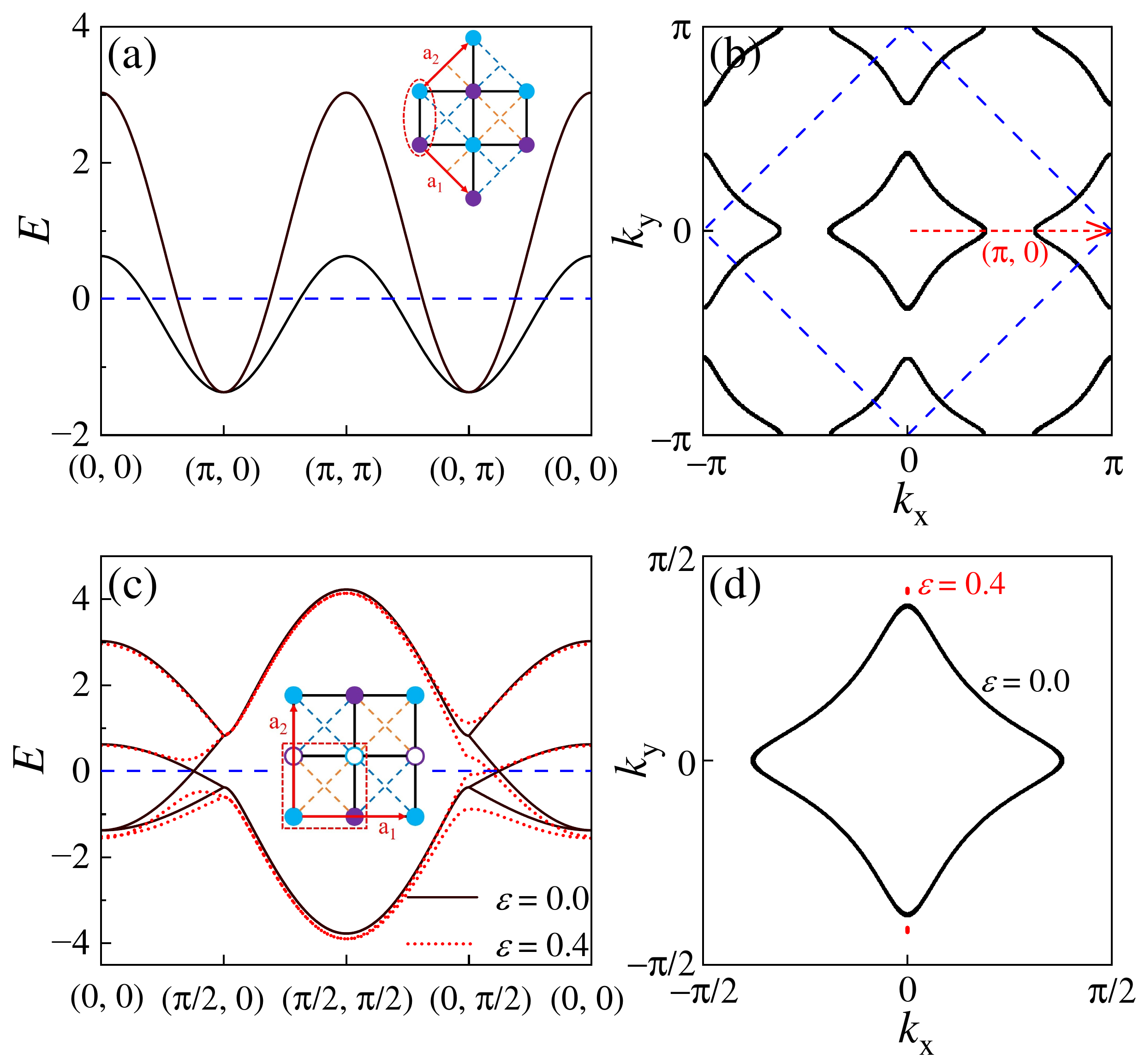}  
\caption{(a) Noninteracting electronic band structure in a two-sublattice basis with hopping elements $t_1=0.3$, $t_2=1.4$, and $t_2'=-0.6$ at half filling. (b) Fermi surface corresponding to the panel (a) with blue dashed line marking the 1st Brillouin zone, which includes the hole pocket at $(0,0)$ and the electron pocket at $(\pi,0)$. (c) Noninteracting electronic band structure in a four-sublattice basis with a uniaxial staggered electric potential $\varepsilon$ at half filling. (d) Fermi surface in the 1st Brillouin zone corresponding to the panel (c), which undergoes a topological transition via introducing $\varepsilon$.}\label{fig_s1}
\end{figure}

Defined on a square lattice with staggered next-nearest-neighbor hoppings, as introduced in the main text, the original nonmagnetic unit cell contains a two-sublattice basis with diagonal primitive lattice vectors, sketched in the inset of Fig.~\ref{fig_s1}(a). The corresponding noninteracting electronic band structure and Fermi surface at half-filling are presented in Figs.~\ref{fig_s1}(a) and \ref{fig_s1}(b), respectively. As expected, the $(0,0)$ hole pocket and $(\pi,0)$ electron pocket emerge, connected by the wavevector $\mathbf{k}=(\pi,0)$, which reproduces the essential electronic starting point of iron pnictides.

In our study, we introduce the charge density inhomogeneity via a uniaxial staggered electric potential $\varepsilon$. Consequently, the original nonmagnetic unit cell is enlarged to a $2\times2$ geometry with expanded primitive lattice vectors, as described in the inset of Fig.~\ref{fig_s1}(c), which naturally induces Brillouin zone folding, as visualized in the folded electronic band [Fig.~\ref{fig_s1}(c)] and Fermi surface [Fig.~\ref{fig_s1}(d)] at half-filling. 
Equivalent to the case shown in Figs.~\ref{fig_s1}(a) and (b), when $\varepsilon=0$, the original electron pocket at $(\pi,0)$ are perfectly folded to the $(0,0)$ point, overlapping with the hole pocket. However, introducing a finite $\varepsilon$ completely reconstructs the Fermi surface topology: the uniaxial symmetry breaking opens a gap along the $(0,0)$-$(\pi/2,0)$ path, while the preserved symmetry in the orthogonal direction protects the gapless Fermi points along the $(0,0)$-$(0,\pi/2)$ path.

\noindent
\section{S2. Self-Consistent Hartree-Fock Equations}

To map the magnetic phase space and evaluate the nonrelativistic spin splitting, we solve the proposed minimal Hubbard model within the Hartree-Fock mean-field framework. The many-body on-site Hubbard interaction is decoupled in the density channel as $U n_{\mathbf{i}l\uparrow} n_{\mathbf{i}l\downarrow} \approx U \langle n_{\mathbf{i}l\uparrow} \rangle n_{\mathbf{i}l\downarrow} + U \langle n_{\mathbf{i}l\downarrow} \rangle n_{\mathbf{i}l\uparrow} - U \langle n_{\mathbf{i}l\uparrow} \rangle \langle n_{\mathbf{i}l\downarrow} \rangle$. 
Using the definition of the spin-stripe order parameter $\delta m_S$ introduced in the main text, the local electron occupation is parameterized as $\langle n_{\mathbf{i}l\sigma} \rangle = \frac{n}{2} + (-1)^{l+\sigma}\delta m_S$ with electron filling $n$, where $\sigma \in \{1, 2\}$ corresponds to spin $\uparrow, \downarrow$. Substituting this into the interaction term, we obtain the effective mean-field interaction:
\begin{align}\label{S_HF_decoupling}
U \sum_{\mathbf{i}l} n_{\mathbf{i}l\uparrow} n_{\mathbf{i}l\downarrow} \approx \sum_{\mathbf{i}l} \bigg[ &U\frac{n}{2} (n_{\mathbf{i}l\uparrow} + n_{\mathbf{i}l\downarrow}) + U \delta m_S (-1)^{l} (n_{\mathbf{i}l\uparrow} - n_{\mathbf{i}l\downarrow}) \bigg] \nonumber \\
&- U \sum_{\mathbf{i}l} \left( \frac{n^2}{4} - \delta m_S^2 \right).
\end{align}
The first term acts as a uniform chemical potential shift, while the second term provides the crucial spin- and sublattice-dependent effective Zeeman field which drives the spin-stripe instability. The last term is a constant energy term, which does not affect the diagonalization process. Focusing on the fixed filling $n$, we only need to consider the effective second term.

After performing the Fourier transformation, the interaction term gives rise to the diagonal matrix elements of the $4\times4$ momentum-space Hamiltonian $H_{\mathbf{k}\sigma}$ defined in Eq.~(4) of the main text. Then the Hartree-Fock equations are solved by self-consistently determining the order parameter, 
\begin{align}\label{S_order_parameter}
\delta m_S &= \frac{1}{8N_c} \sum_{\mathbf{i}} \langle S^z_{\mathbf{i}1} - S^z_{\mathbf{i}2} + S^z_{\mathbf{i}3} - S^z_{\mathbf{i}4} \rangle_{\mathrm{HF}} \nonumber \\
&= \frac{1}{8N_c} \sum_{\mathbf{k} \alpha \sigma} \Psi_{\mathbf{k} \alpha \sigma}^\dagger \mathcal{M}_S \Psi_{\mathbf{k} \alpha \sigma} f(E_{\mathbf{k} \alpha \sigma} - \mu),
\end{align}
where $E_{\mathbf{k} \alpha \sigma}$ and $\Psi_{\mathbf{k} \alpha \sigma}$ are the $\alpha$-th eigenvalues and corresponding eigenvectors with spin $\sigma$ of the total Hamiltonian $H^{\mathrm{HF}}_\mathbf{k}=\text{diag}(H_{{\bf k}\uparrow}, H_{{\bf k}\downarrow})$, and the diagonal $\mathcal{M}_S = \text{diag}(1, -1, 1, -1, -1, 1, -1, 1)$ is the signature matrix defining the spin-stripe configuration. Simultaneously, the chemical potential $\mu$ must be dynamically adjusted to strictly enforce the half-filling condition:
\begin{equation}\label{S_density}
n = \frac{1}{4N_c} \sum_{\mathbf{k} \alpha \sigma} f(E_{\mathbf{k} \alpha \sigma} - \mu) = 1,
\end{equation}
where $f(E_{\mathbf{k} \alpha \sigma} - \mu)$ is the Fermi-Dirac distribution function at $T = 0$ for the ground state.

For the numerical evaluation, we discretize the 1st Brillouin zone ($k_x, k_y\in(-\pi/2,\pi/2]$) using a dense mesh of $200 \times 200$ momentum points. The self-consistent equations are solved iteratively. In the iteration, we construct and diagonalize the matrix $H^{\mathrm{HF}}_\mathbf{k}$ using an input guess $\delta m_S^{\text{in}}$. We then determine $\mu$ by satisfying Eq.~\eqref{S_density}, and subsequently compute the updated order parameter $\delta m_S^{\text{out}}$ using Eq.~\eqref{S_order_parameter} as input for the next iteration. To improve convergence and suppress oscillatory divergence of order parameters, we employ a linear mixing scheme by updating the input for the next step as $\delta m_S^{\text{next}} = w \delta m_S^{\text{out}} + (1-w) \delta m_S^{\text{in}}$ with a mixing weight $w$. This iterative procedure is strictly repeated until the order parameter converges to the accuracy of $|\delta m_S^{\text{out}} - \delta m_S^{\text{in}}| < 10^{-6}$. Moreover, the N\'eel antiferromagnetic ($\delta m_N$) and total ferromagnetic ($\delta m_T$) order parameters follow the same analysis and iteration strategy.

\noindent
\section{S3. Determinant quantum Monte Carlo method and sign problems}
In the main text, we simulate our model using the unbiased determinant quantum Monte Carlo (DQMC) method. In the DQMC algorithm, the observable at finite temperatures is evaluated in the grand canonical ensemble as: $\langle O \rangle = \operatorname{Tr} (\mathrm{e}^{-\beta H} O) / \operatorname{Tr} (\mathrm{e}^{-\beta H})$. In simulations, the $\mathrm{e}^{-\beta H}$ is discretized into small time slices $e^{-\beta H} = \prod_L e^{-\Delta \tau H}$ with $L\Delta \tau=\beta$ via the Trotter decomposition. In our study, we use a sufficiently small $\Delta \tau=0.1$, so that the Trotter errors $\mathcal{O}(\Delta \tau^2)$ are smaller than those associated with the statistical sampling.
Subsequently, to decouple the four-fermion interaction term $U$, we employ the discrete Hubbard-Stratonovich (HS) transformation to couple the fermions to the spin channel:
\begin{equation}\label{S_channel}
e^{- \Delta\tau U \sum_{{\bf i}l} n_{{\bf i}l\uparrow}n_{{\bf i}l\downarrow}} \rightarrow e^{ \Delta\tau \frac{U}{2} \sum_{{\bf i}l} (n_{{\bf i}l\uparrow}-n_{{\bf i}l\downarrow})^2} = \frac{1}{4} \prod_{{\bf i}l} \sum_{\alpha=\pm 1, \pm 2} \gamma_{{\bf i}l}(\alpha) e^{\sqrt{\Delta\tau\frac{U}{2}}\eta_{{\bf i}l}(\alpha)(n_{{\bf i}l\uparrow}-n_{{\bf i}l\downarrow})} + \mathcal{O}((\Delta\tau\frac{U}{2})^4),
\end{equation}
where the standard coefficients of HS are defined as $\gamma(\pm 1) = 1 + \sqrt{6}/3, \gamma(\pm 2) = 1 - \sqrt{6}/3$ and $\eta(\pm 1) = \pm \sqrt{2(3 - \sqrt{6})}, \eta(\pm 2) = \pm \sqrt{2(3 + \sqrt{6})}$. After HS transformation, through integrating out the fermionic degrees of freedom, the partition function can be written as
\begin{align}\label{partition}
Z = \operatorname{Tr} \left( e^{-\beta H} \right) &= \sum_{\{s\}} \left( \prod_{{\bf i}l} \prod_{\ell=1}^{L} \gamma_{{\bf i}l,\ell}(\{s\}) \right) \times \prod_{\sigma=\uparrow,\downarrow} \det \left( 1 + B_{L}^{\sigma}(\{s\})B_{L-1}^{\sigma}(\{s\}) \cdots B_{1}^{\sigma}(\{s\}) \right)  \\
&= \sum_{\{s\}} P(\{s\}),
\end{align}
where $B_{\ell}^{\sigma} = e^{-\Delta\tau \mathbf{K}^{\sigma}} \prod_{{\bf i}l} e^{\sqrt{\Delta\tau \frac{U}{2}} \eta_{{\bf i}l,\ell} (\{s\}) \mathbf{V}^{\sigma}_{{\bf i}l}}$ for a given auxiliary field configuration $\{s\}$. Therefore, the partition function is expressed as a sum over the auxiliary-field weights $P$ (i.e., the fermion determinant) and the observable becomes $\langle O \rangle=\frac{\sum_{\{s\}}P(\{s\})O(\{s\})}{Z}$, based on sampling over the auxiliary fields.

In the DQMC algorithm, the sign problem is a major limitation. The sign problem arises when the determinant $P(\{s\})$ becomes negative. We take $P(s)$ as the real part of the fermion determinant by default, and when the average sign is not too small, the observable can be evaluated within a reweighting scheme. By defining $S(\{s\})=\frac{P(\{s\})}{|P(\{s\})|}$, $\langle O \rangle= \frac{\sum_{\{s\}} P(\{s\})O(\{s\}) / \sum_{\{s\}} |P(\{s\})|}{\sum_{\{s\}} P(\{s\}) / \sum_{\{s\}} |P(\{s\})|}=\frac{\langle S O \rangle_{\{s\}}}{\langle S \rangle_{\{s\}}}$, where the average sign $\langle S \rangle_{\{s\}}= \frac{\sum_{\{s\}} |P(\{s\})|S(\{s\})}  {\sum_{\{s\}} |P(\{s\})|}$. For more technical details about DQMC, please see Refs.~\cite{PhysRevB.40.506,Assaad2008}.

\begin{figure}[h]
\centering
\includegraphics[width=0.7\textwidth]{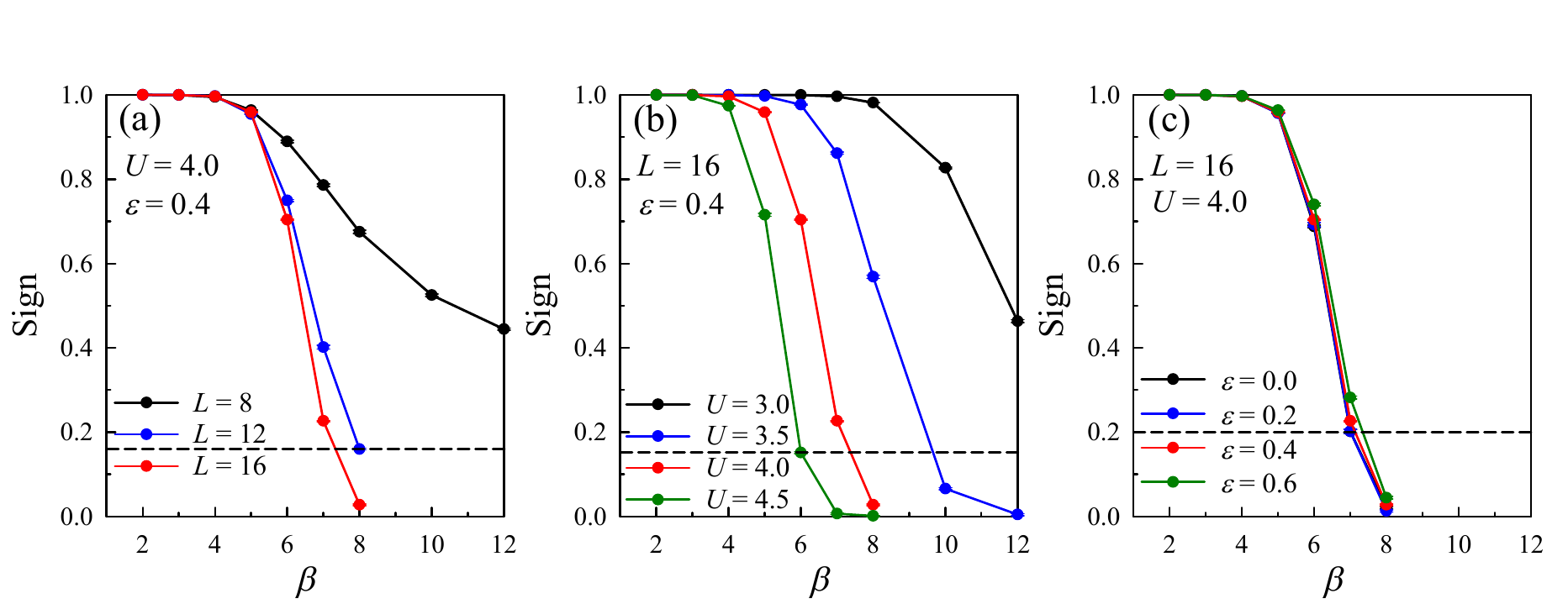}  
\caption{Average signs as a function of inverse temperature $\beta$ for (a) diverse $L$ at a fixed $U=4.0, \varepsilon=0.4$, (b) diverse $U$ at a fixed $L=16, \varepsilon=0.4$, and (c) diverse $\varepsilon$ at a fixed $L=16, U=4.0$. The sign is significantly suppressed by $\beta$, $L$, and $U$, while it is slightly promoted by $\varepsilon$. The dashed line marks the minimum average sign level involved in the main text.}\label{fig_s2}
\end{figure}

In bipartite lattices, such as square lattice or honeycomb lattice considering the nearest-neighbor hopping, the half-filled system exhibits no sign problem due to particle-hole symmetry. However, we introduce alternating next-nearest-neighbor hopping on our square lattice, which breaks particle-hole symmetry and thus gives rise to a sign problem. The average sign in our system is presented in Fig.~\ref{fig_s2}.

As expected, the average sign is reduced with temperature decreasing and size increasing in Fig.~\ref{fig_s2}(a). Increasing the on-site interaction $U$ severely suppresses the average sign, as shown in Fig.~\ref{fig_s2}(b). In Fig.~\ref{fig_s2}(c), the introduction of uniaxial staggered electric potential $\varepsilon$ slightly enhances the average sign, although the effect is negligible. In our study, a temperature of $\beta=6$ is mainly adopted, which captures a good average sign for all system sizes $L$ with all $U$ and $\varepsilon$ parameters. This provides compelling numerical evidence for our study.

\noindent
\section{S4. The spin susceptibility}
\begin{figure}[h]
\centering
\includegraphics[width=0.7\textwidth]{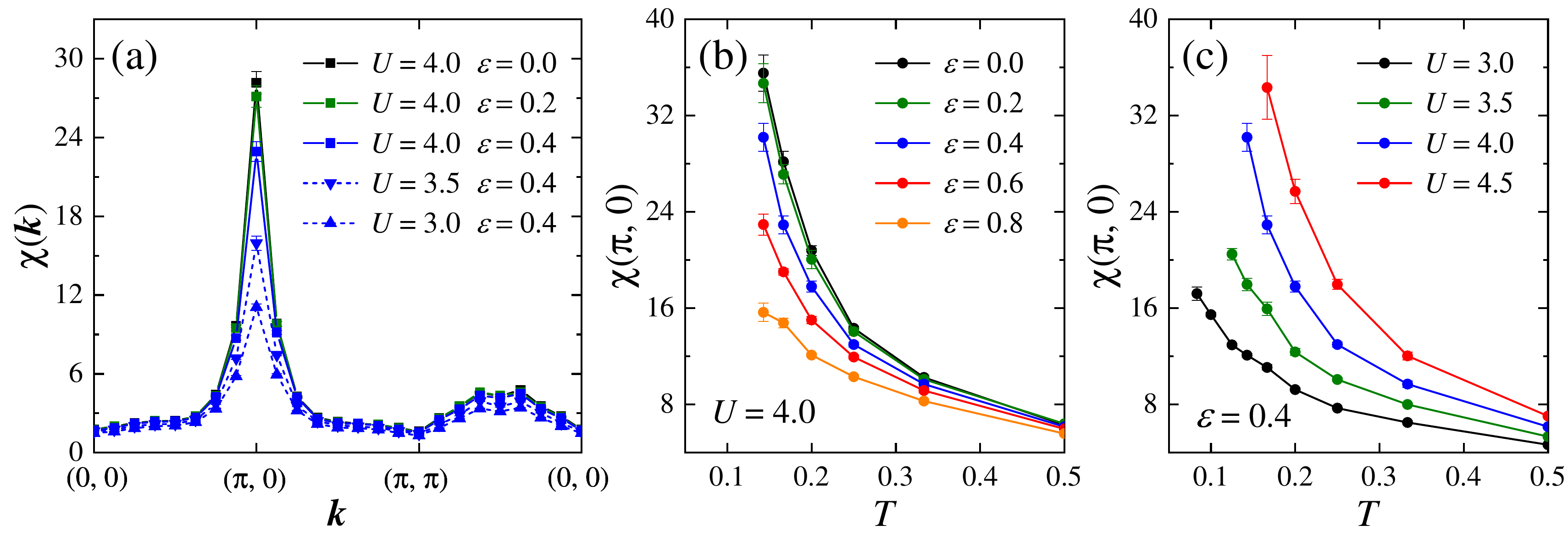}  
\caption{Spin susceptibility $\chi$. (a) Momentum-dependent $\chi({\mathbf{k}})$ along the high-symmetry path of the unfolded Brillouin zone for different $\varepsilon$ and $U$ at $\beta=6$, which indicates a highly dominant $(\pi,0)$ spin-density-wave instability. (b), (c) Temperature $T$-dependence of the spin-stripe susceptibility $\chi(\pi,0)$ for (b) diverse $\varepsilon$ at a fixed $U=4.0$ and (c) diverse $U$ at a fixed $\varepsilon=0.4$, where divergence of $\chi(\pi,0)$ at low temperatures signals the stabilization of the spin-stripe order. All data are obtained from the system size $L=16$.}\label{fig_s3}
\end{figure}

In DQMC simulations, we also measure the imaginary-time spin susceptibility of the system, defined as
\begin{equation}\label{chi}
\chi(\mathbf{k}) = \int_0^\beta d\tau \frac{1}{N_c} \sum_{\mathbf{i}l, \mathbf{j}l'} e^{-i\mathbf{k}\cdot(\mathbf{R}_{\mathbf{i}l} - \mathbf{R}_{\mathbf{j}l'})} \langle S^z_{\mathbf{i}l}(\tau) S^z_{\mathbf{j}l'}(0) \rangle,
\end{equation}
to capture the low-energy dynamical magnetic fluctuations, as shown in Fig.~\ref{fig_s3}.

Firstly, we evaluate the momentum-resolved response of the spin susceptibility $\chi$, as presented in Fig.~\ref{fig_s3}(a). A remarkably sharp peak emerges at the wavevector $\mathbf{k}=(\pi,0)$, signaling a dominant $(\pi,0)$ spin-density-wave instability. This behavior is in excellent agreement with results of spin structure factors. 
According to the general theory of phase transitions, the susceptibility of the corresponding order parameter diverges as temperature $T$ approaches the critical point. Therefore, tracing the $T$-dependence of the spin-stripe susceptibility $\chi(\pi,0)$ is an intriguing signature for the stripe order, as illustrated in Figs.~\ref{fig_s3}(b) and \ref{fig_s3}(c). 

Focusing on a representative electron correlation $U=4.0$ within the stripe-ordered altermagnetic phase, although increasing the staggered potential $\varepsilon$ suppresses the low-$T$ divergence of $\chi(\pi,0)$, a steep diverging tendency survives at finite $\varepsilon$ (e.g., $\varepsilon=0.0$, $0.2$, and $0.4$), positively suggesting the thermodynamic stabilization of the underlying spin-stripe configuration. For a fixed $\varepsilon=0.4$, Increasing $U$ dramatically enhances this divergent trend, profoundly underscoring the correlation-driven stripe-ordered altermagnetism. 
These data are obtained within parameters with tractable sign problems, where exact numerical results of $\chi$ further supports our conclusions.

\end{document}